# Simultaneous Presentation of Thermal and Mechanical Stimulation Using High-Intensity Airborne Ultrasound

Sota Iwabuchi[1], Ryoya Onishi[1], Shun Suzuki[1], Takaaki Kamigaki[1], Yasutoshi Makino[1], and Hiroyuki Shinoda[1]

[1] *Graduate School of Frontier Sciences, The University of Tokyo, Japan*

*(Email:iwabuchi@hapis.k.u-tokyo.ac.jp)*

**Abstract ---** In this study, we propose a non-contact thermal presentation method using airborne ultrasound. We generate strong sound field directly on the human skin and present a perceivable temperature rise. The proposed method enables simultaneous presentation of mechanical and thermal stimuli. In preliminary experiments, we confirmed that temperature increase of 5.4 °C occurs at the palm after 5.0 s.

## 1 INTRODUCTION

With the increasing popularity of virtual reality (VR), research on haptic displays that reproduce real tactile information has been attracting attention. Among these, displays that provide a thermal sensation are particularly important for expressing the temperature of the virtual space environment and the temperature information of virtual objects. By presenting thermal stimuli without directly touching the device, it is possible to provide a more natural and immersive VR experience that does not limit the user's movements.

In this study, we propose a non-contact thermal sensation presentation method using airborne ultrasound. By focusing powerful ultrasound on the skin, we can simultaneously provide mechanical pressure and temperature stimuli. The proposed method is invisible to the human eye, unlike conventional methods that use visible laser light or halogen lamps. Due to this characteristic, we believe that this research is effective in applications that interfere with the real world, such as AR.

## 2 PROTOTYPE

### 2.1 Overview

Airborne ultrasound tactile display is a haptic device that can present pressure to the user without physical contact using sound wave interference.

It has been known that the temperature at the skin surface increases when strong acoustic energy is supplied by airborne ultrasound [1] [2]. In this study, we propose a method of simultaneous presentation of mechanical and thermal stimuli based on this phenomenon.

### 2.2 System Configuration

Figure 1 shows a photograph of the temperature presentation system that will be used in the demonstration. This system consists of 12 ultrasound phased-array devices. We used AUTD3 [3] for the phased array hardware and software. A thermography camera (OPTPI 45ILTO29T090, Optris) is attached to the center of the phased array to record skin temperatures.

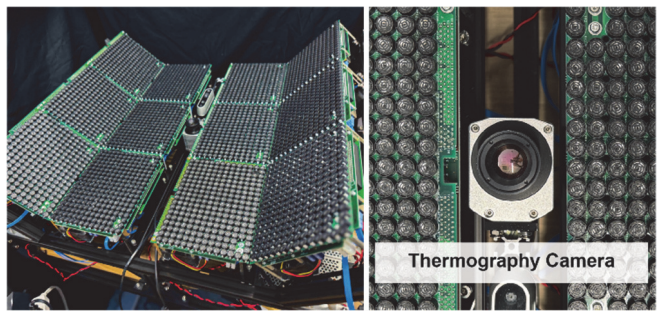

Fig.1 Photograph of the experimental system: 12 phased arrays with a thermal imaging camera mounted in the center to obtain skin temperatures.

### 2.3 Design of Sound Field Conditions

In the demonstration, participants will experience two types of tactile stimulation using ultrasound. The first is static focused ultrasound with the amplitude fixed at its maximum value, and the second is focused ultrasound with square wave amplitude modulation. The duty ratio of the square wave, i.e., the ratio of the time the ultrasound is on, is set to 0.9 to provide sufficient acoustic energy.

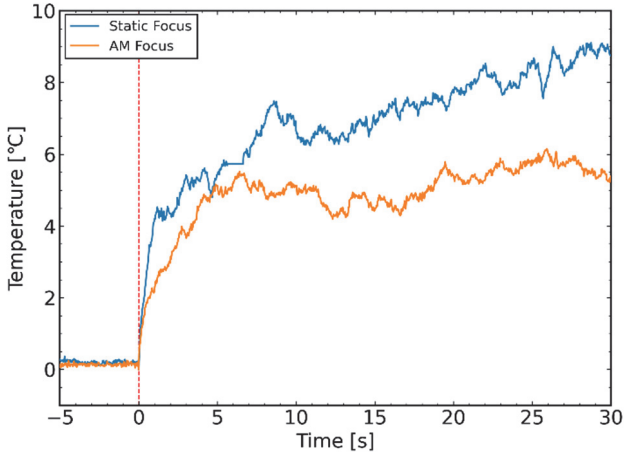

Fig.2 Time variation of skin temperature at the irradiated position. The horizontal axis represents time, and the vertical axis represents temperature change from the temperature at the start of irradiation.

### 2.4 Preliminary Evaluation

We conducted a preliminary experiment to evaluate the performance of the thermal presentation. Two stimuli introduced in subsection 2.3 were presented to the center of the palm at a distance of 296 mm from the six phased arrays in the center. The temperature of the skin surface was measured by a thermography camera.

As shown in Fig.2, the experimental results showed that the static stimulus produced a temperature increase of 5.4 °C in 5.0 s, and the vibratory stimulus, using a 50 Hz square wave amplitude modulation, induced a temperature increase of 4.5 °C in 5.0 s. After 30 s, we observed a temperature increase of 8.6 °C in the case of the static stimulus and a temperature increase of 5.4 °C in the case of the vibration stimulus (Fig. 3).

According to previous study [4], when the acclimation temperature of the palm is 33 °C, the threshold for feeling thermal sensation is 0.2 °C. The results of the preliminary experiments exceeded this threshold, and we believe that the performance of the device is sufficient for use in the presentation of a thermal sensation.

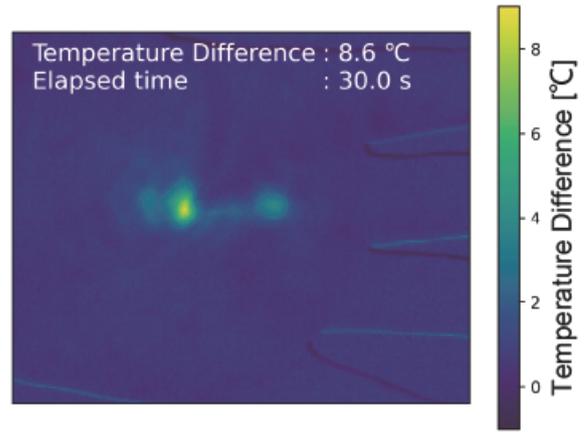

Fig.3 Distribution of temperature variation of static focus. This image was taken 30.0s after the ultrasound irradiation.

## 3 Conclusion

In this paper, we proposed a method for non-contact presentation of mechanical and thermal stimuli using airborne ultrasound. In the future, we would like to investigate the psychological changes that occur when two stimuli are presented simultaneously.

### Acknowledgement

This work was supported by JST Moonshot R&D Program Grant Number JPMJMS239E-01 and JST SPRING, Grant Number JPMJSP2108.